\documentclass[onecolumn,unpublished,a4paper]{quantumarticle}
\usepackage[utf8]{inputenc}
\usepackage[backend=biber,style=alphabetic,maxbibnames=999]{biblatex}
\usepackage{amsmath}
\usepackage{amssymb}
\usepackage{amsthm}
\usepackage{amsfonts}
\usepackage[caption=false]{subfig}
\usepackage[colorlinks]{hyperref}
\usepackage[all]{hypcap}
\usepackage{tikz}
\usepackage{relsize}
\usepackage{color,soul}
\usepackage{capt-of}
\usepackage{mathtools}
\usepackage{float}
\usepackage[section]{placeins}
\usepackage{listings}
\usepackage[T1]{fontenc}  
\usetikzlibrary{decorations.pathreplacing}
\usepackage{braket}

\DeclareFixedFont{\ttb}{T1}{txtt}{bx}{n}{4}
\DeclareFixedFont{\ttm}{T1}{txtt}{m}{n}{4}
\definecolor{deepblue}{rgb}{0,0,0.5}
\definecolor{deepred}{rgb}{0.6,0,0}
\definecolor{deepgreen}{rgb}{0,0.5,0}
\newcommand\cppstyle{\lstset{
language=C++,
basicstyle=\ttm,
otherkeywords={uint8_t, __m256i, size_t, ASSERT_TRUE, EXPECT_TRUE, TEST, BENCHMARK},
keywordstyle=\ttb\color{deepblue},
emphstyle=\ttb\color{deepblue},
stringstyle=\color{deepgreen},
commentstyle=\fontfamily{txtt}\selectfont\color{gray},
showstringspaces=false,
literate={*}{{\char42}}1
         {-}{{\char45}}1
}}
\lstnewenvironment{cpp}[1][]
{\cppstyle\lstset{#1}}{}

\newcommand\pythonstyle{\lstset{
language=python,
basicstyle=\ttm,
morekeywords={assert,as,echo},
keywordstyle=\ttb\color{deepblue},
emphstyle=\ttb\color{deepblue},
stringstyle=\color{deepgreen},
commentstyle=\fontfamily{txtt}\selectfont\color{gray},
showstringspaces=false,
literate={*}{{\char42}}1
         {-}{{\char45}}1
}}
\lstnewenvironment{python}[1][]
{\pythonstyle\lstset{#1}}{}

\lstdefinestyle{stimcircuit}{
    language=python,
    basicstyle=\fontsize{6}{6}\selectfont\ttfamily,
    upquote=true,
    stepnumber=1,
    numbersep=8pt,
    showstringspaces=false,
    breaklines=true,
    frame=single,
    aboveskip=1.5em,
    belowskip=1.5em,
    commentstyle=\color{gray},
    classoffset=1,
    morekeywords={DETECTOR,OBSERVABLE_INCLUDE,rec},
    keywordstyle=\color{deepgreen},
    classoffset=2,
    morekeywords={H,R,MPP,M,RX,RY,MY,MX,SQRT\_X,XCY,XCZ,YCX},
    keywordstyle=\color{deepblue},
    classoffset=3,
    morekeywords={X_ERROR,DEPOLARIZE2,DEPOLARIZE1},
    keywordstyle=\color{red},
    classoffset=4,
    morekeywords={TICK,SHIFT_COORDS,QUBIT_COORDS},
    keywordstyle=\color{gray}
}

\renewcommand\thesection{\arabic{section}}

\theoremstyle{definition}

\theoremstyle{definition}

\theoremstyle{definition}

\newcommand{\eq}[1]{\hyperref[eq:#1]{Equation~\ref*{eq:#1}}}
\renewcommand{\sec}[1]{\hyperref[sec:#1]{Section~\ref*{sec:#1}}}
\DeclareRobustCommand{\app}[1]{\hyperref[app:#1]{Appendix~\ref*{app:#1}}}
\newcommand{\fig}[1]{\hyperref[fig:#1]{Figure~\ref*{fig:#1}}}
\newcommand{\tbl}[1]{\hyperref[tbl:#1]{Table~\ref*{tbl:#1}}}
\newcommand{\theoremref}[1]{\hyperref[theorem:#1]{Theorem~\ref*{theorem:#1}}}
\newcommand{\definitionref}[1]{\hyperref[definition:#1]{Definition~\ref*{definition:#1}}}
\newcommand{\propref}[1]{\hyperref[property:#1]{Property~\ref*{property:#1}}}

\begin{document}
\title{Less Bacon More Threshold}

\date{\today}

\author{Craig Gidney}
\email{craig.gidney@gmail.com}
\affiliation{Google Quantum AI, Santa Barbara, California 93117, USA}

\author{Dave Bacon}
\affiliation{Google Quantum AI, Seattle, Washington 98103, USA}

\maketitle

\begin{abstract}
We give the Bacon-Shor code a threshold purely by deleting gates from its circuit.
Specifically: we use lattice surgery to concatenate the Bacon-Shor code with itself using local planar connectivity, and observe that the resulting circuit is a subset of the circuit that would be used by a larger Bacon-Shor code.
\end{abstract}

\emph{The code written, circuits generated, and stats collected for this paper are available on Zenodo at \href{https://doi.org/10.5281/zenodo.7901729}{doi.org/10.5281/zenodo.7901729}~\cite{gidneybacondata}.}

\section{Introduction}
\label{sec:introduction}

The Bacon-Shor code~\cite{bacon2006operator} is a quantum error correcting code built out of two-qubit parity measurements (``pair measurements'') on a planar grid of qubits.
Each vertical edge of the grid corresponds to a $Z \otimes Z$ measurement.
Each horizontal edge corresponds to an $X \otimes X$ measurement.
The code's logical qubit is defined by the anticommuting observables formed by the product of $X$ operators along the top row versus the product of $Z$ operators along the left column.
The stabilizer generators of the code are the products of adjacent rows of $X$ operators, and adjacent columns of $Z$ operators.

The Bacon-Shor code was initially considered for its thermodynamic properties, but then attracted considerable attention due to its simplicity\cite{aliferis2007subsystem} and amenability to hardware.
It has found applications in biased noise proposals~\cite{li2019compasslerpsurface, huang2020fault, napp2012optimal, brooks2013fault} and has been realized in experiment~\cite{li2018direct, egan2021fault}.
However, the Bacon-Shor code has a major problem: it doesn't scale to arbitrarily low error rates.
It has no threshold.
For a fixed physical error rate, increasing the size of the code initially improves the logical error rate, but this improvement eventually stops and reverses.
The underlying issue is that the stabilizers being compared by the code all run along the entire length of the grid.
This length is increasing with the size of the code.
As the size increases, forming the stabilizers involves combining more and more measurements.
This makes the stabilizers noisier and noisier.
Eventually, the stabilizer noise overwhelms the gains in code distance, and performance regresses.

The lack of threshold in the Bacon-Shor code can be fixed in a variety of ways.
For example, it can be fixed by mixing in small patches of surface code~\cite{li2019compasslerpsurface} or by concatenation of the code with itself~\cite{aliferis2007subsystem, cross2007comparative}.
Our work focuses on obtaining a threshold using the simplest strategy possible - subtracting gates - requiring no additional connectivity or circuit overhead.
The key underlying idea is that we can concatenate the Bacon-Shor code with itself by using lattice surgery, and the gates used during the lattice surgery are a subset of the gates that would be used by a non-concatenated Bacon-Shor code.  This concatenation differs from the concatenation in~\cite{aliferis2007subsystem, cross2007comparative} because those constructions require long range connectivity to implement higher levels of concatenation.

The paper proceeds as follows.
First, \sec{construction} describes the lattice surgery pair measurement, the equivalence between concatenation and gate deletion, and the pattern of gates we deleted.
Second, \sec{benchmark} benchmarks the construction, showing that it outperforms the normal Bacon-Shor code.
Finally, \sec{conclusion} gives some closing remarks.

\section{Construction}
\label{sec:construction}

\begin{figure}[H]
    \centering
    \resizebox{\linewidth}{!}{
    \includegraphics{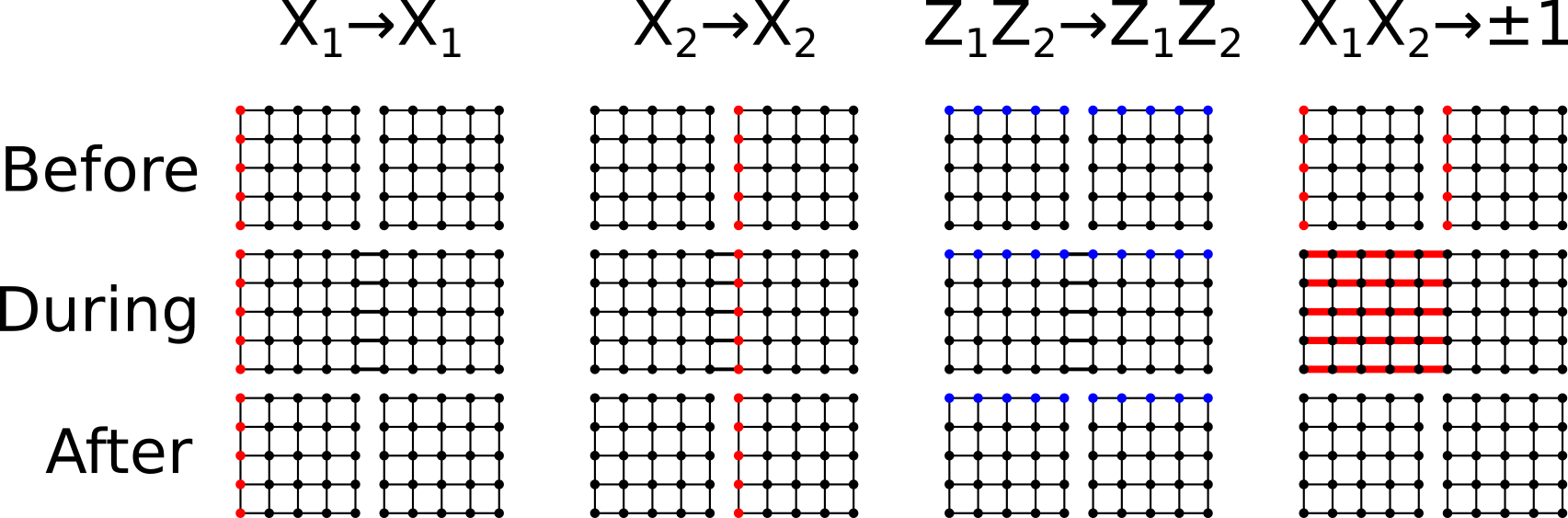}
    }
    \caption{
        Lattice surgery in the Bacon-Shor code.
        When two Bacon-Shor logical qubits are next to each other, temporarily activating the pair measurements between them performs a logical pair measurement.
        This implementation satisfies all the rules required of a logical $XX$ pair measurement~\cite{mcewen2023relaxingsurface}: it preserves the individual logical X observables, preserves the logical $ZZ$ parity, and reveals the logical $XX$ parity (via the parity of the measurements of all of the edges between the two logical observables).
        In the diagram, drawn vertical edges are active physical $ZZ$ measurements.
        Drawn horizontal edges are active physical $XX$ measurements.
        Nodes colored red indicate the support of a logical $X$.
        Nodes colored blue indicate the support of the logical $ZZ$.
        Edges colored red indicate measurements revealing the logical $XX$.
    }
    \label{fig:lattice_surgery}
\end{figure}

Lattice surgery is a well established technique for performing pair measurements in the surface code~\cite{de2017zxlattice,fowler2018latticesurgery}.
Lattice surgery also works on the Bacon-Shor code (see \fig{lattice_surgery})~\cite{poulsen2017lattice}.
Placing two Bacon-Shor codes next to each other, and temporarily merging them into one larger code, performs a logical pair measurement.

An important distinction between surface code lattice surgery, and Bacon-Shor lattice surgery, is that Bacon-Shor lattice surgery incentivizes monogamy.
In surface code lattice surgery, we want to involve every logical qubit in as many simultaneous surgeries as possible, because this makes the computation denser.
In Bacon-Shor lattice surgery we would still like to make the computation as dense as possible, but there's a problem: the stabilizers running between the logical qubits are longer during a surgery.
For example, if we simultaneously stitched a whole column of logical qubits, the stabilizers could get so long and noisy that the noise would become overwhelming and break fault tolerance.
A simple way to avoid this catastrophe is to only involve each logical Bacon-Shor qubit in one surgery at a time.

Lattice surgery gives us a logical pair measurement that only requires local planar connectivity.
As long as there is a code distance where the logical pair measurement outperforms the physical pair measurements, arbitrarily good logical qubits can be created by concatenation.
Therefore, by using lattice surgery to concatenate the Bacon-Shor code with itself, we have created a fault tolerant construction with a threshold.
See \app{threshold} for a more detailed proof.

If we produce a circuit implementing a Bacon-Shor code concatenated with itself using lattice surgery, we observe something interesting: the circuit is using a strict subset of the operations that would have been used by a non-concatenated Bacon-Shor code covering the same set of physical qubits.
This is because the logical pair measurements are activated by including physical pair measurements between the logical qubits.
In a non-concatenated Bacon-Shor code, these physical pair measurements would have simply always been included, instead of only being included while performing the logical pair measurement.
Therefore we can switch our perspective and view our concatenated code construction as a construction that deletes gates from a Bacon-Shor circuit.

To use the edge-deletion perspective, we have to specify the pattern of deletions.
For this paper, we found that the following pattern works well:

$$\text{Include}_{b,f}(e, t) = \left(t \equiv b \cdot \frac{4^{L(e, f) + 1}-1}{3} \pmod {4^{L(e, f) + 1}}\right)$$

\begin{figure}[H]
    \centering
    \resizebox{0.9\linewidth}{!}{
    \includegraphics{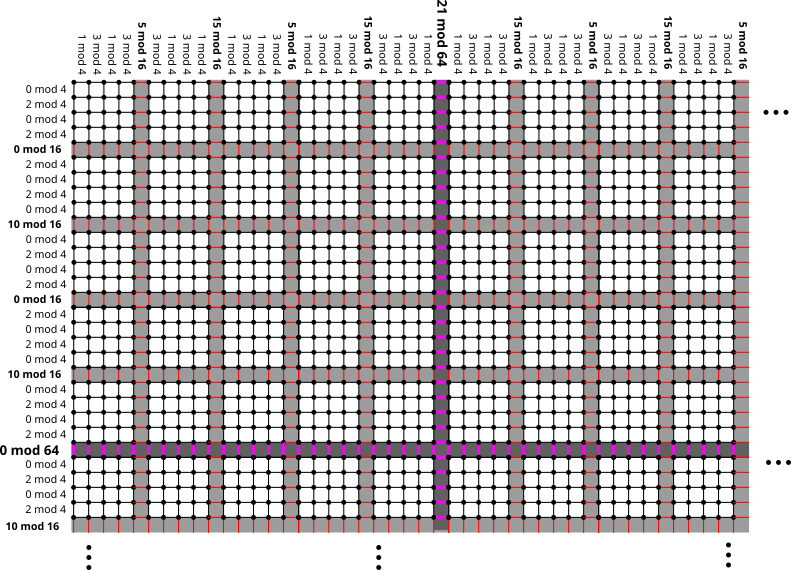}
    }
    \caption{
        The pattern of pair measurements over time, assuming a fractal pitch of 5.
        In each layer of the circuit, pair measurements are included (or deleted) depending on the annotated conditions for that edge's row or column.
        For example, the vertical pair measurements in the row to the right of the label $10 \bmod 16$ are only included in circuit layers where the layer index $t$ satisfies $t \equiv 10 \pmod{16}$.
        Highlights show the underlying concatenated structure.
    }
    \label{fig:cut_pattern}
\end{figure}

In the above equation: $t$ is a circuit layer index, $e$ is the row index of vertical edges or column index of horizontal edges, $L(e,f)$ is the number of times $f$ divides into $e$, $f$ is the ``fractal pitch'' parameter that determines the code distance used when concatenating, and $b$ is an interleaving control parameter set to 0 for odd-index rows and 2 for even-odd rows and 1 for odd-index columns and 3 for even-index columns.
The different $b$ values ensure qubits are only involved in one pair measurement at a time, at all levels of concatenation, assuming an odd fractal pitch.
For each edge, at each time step, we evaluate $\text{Include}_{b,f}(e, t)$ and delete the edge from that layer if the function evaluates to false.
An example pattern of removed measurements is shown in \fig{cut_pattern}.

\begin{figure}[H]
    \centering
    \resizebox{0.7\linewidth}{!}{
    \includegraphics{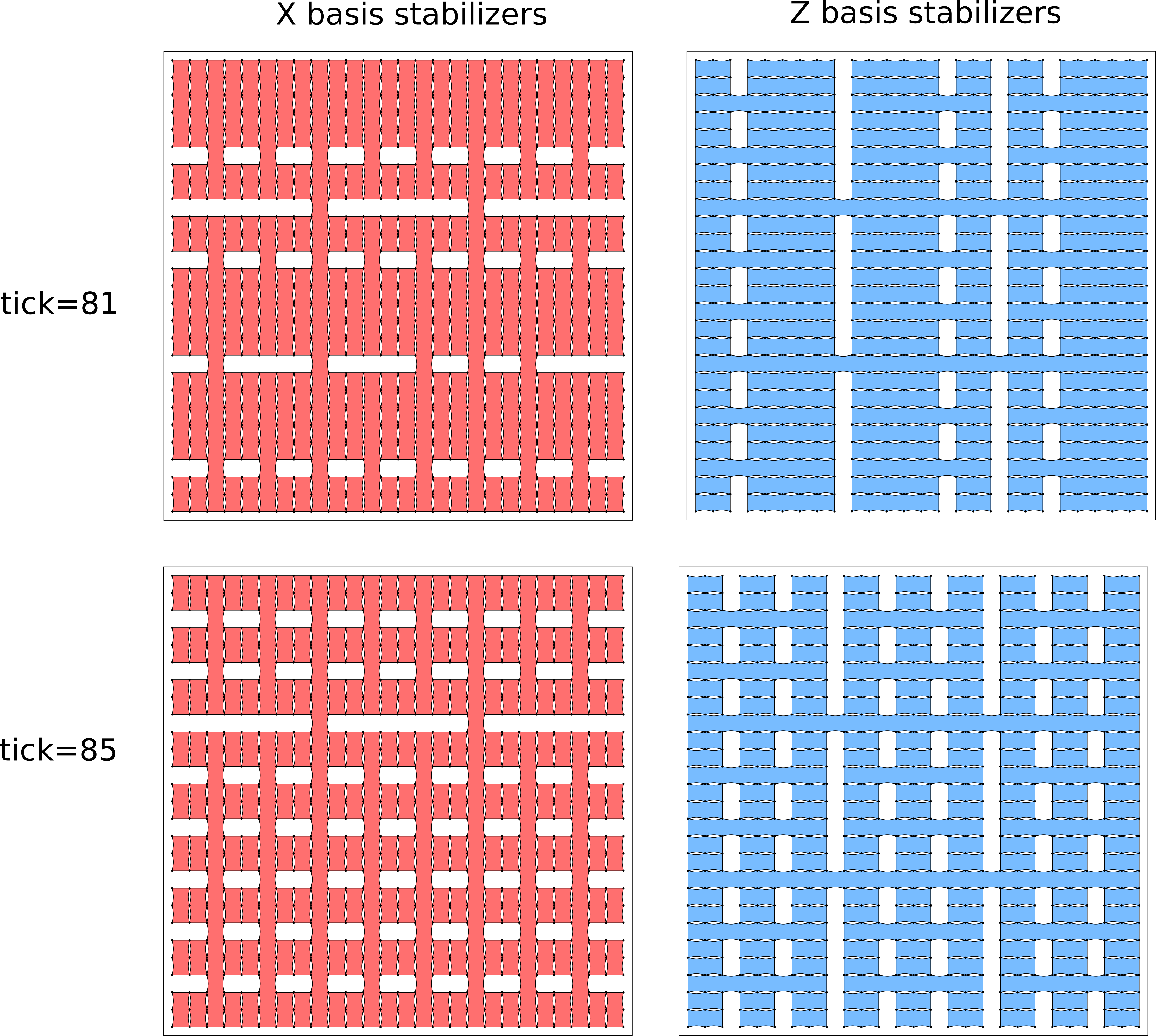}
    }
    \caption{
        Detector slice diagrams showing checked stabilizers of our construction, with a grid diameter of 27 and a fractal pitch of 3.
        The coming and going of the pair measurements at higher levels of concatenation causes variation over time.
        The checked stabilizers correspond to different codes at different time steps.
    }
    \label{fig:detslice}
\end{figure}

Given the pattern of edge inclusions, a maximal set of detectors can be computed using a union find data structure.
When a column of horizontal edges (or row of vertical edges) is measured, initialize a union find data structure with each edge in its own set.
Then, for each anticommuting edge that was included in any layer between the current layer and the last time the column (or row) was measured, perform a union between the two edges the anticommuting edge is touching.
After this is complete, each disjoint set of edges in the data structure corresponds to a detector in the circuit.
Given a disjoint set, its detector is formed by multiplying together the pair measurements of each edge in the set from the current layer and from the previous layer where this column or row was measured.
Taking time slices of the resulting detectors reveals a fractal pattern of checked stabilizers, which varies over time, as shown in \fig{detslice}.
Note that there are substantially more stabilizers being checked than there would be in a normal Bacon-Shor code, which is one reason to expect better performance.

Importantly, although the pattern of stabilizers is more complex than the original pattern of Bacon-Shor stabilizers, they still have the property that errors in the circuit are graphlike (producing an even number of $X$-type and $Z$-type detection events in the bulk).
The circuit can still be decoded by using matching.
In fact, using existing open source tools, this process is entirely automated.
Starting from a circuit annotated with the detectors and observables, Stim~\cite{gidney2021stim} can sample from this circuit and convert it into a decoding graph.
Then, PyMatching~\cite{higgott2021pymatching,higgott2023sparseblossom} can use the decoding graph to predict whether the logical observables were flipped or not, given the sampled detection event data.

\section{Benchmarking}
\label{sec:benchmark}

We compared our construction to the normal Bacon-Shor using Monte-Carlo sampling backed by Stim~\cite{gidney2021stim} and PyMatching~\cite{higgott2021pymatching,higgott2023sparseblossom}.
We built the circuits out of a purely dissipative gate set, with the allowed gates being pair measurements ($M_{XX}$ and $M_{ZZ}$), measurements ($M_X$ and $M_Z$), and resets ($R_{X}$ and $R_{Z}$).
We used a uniform noise model specified in \app{noise_model}.
For each circuit we sampled 100 million shots or 1000 errors, whichever came first.

\begin{figure}[H]
    \centering
    \resizebox{0.6\linewidth}{!}{
    \includegraphics{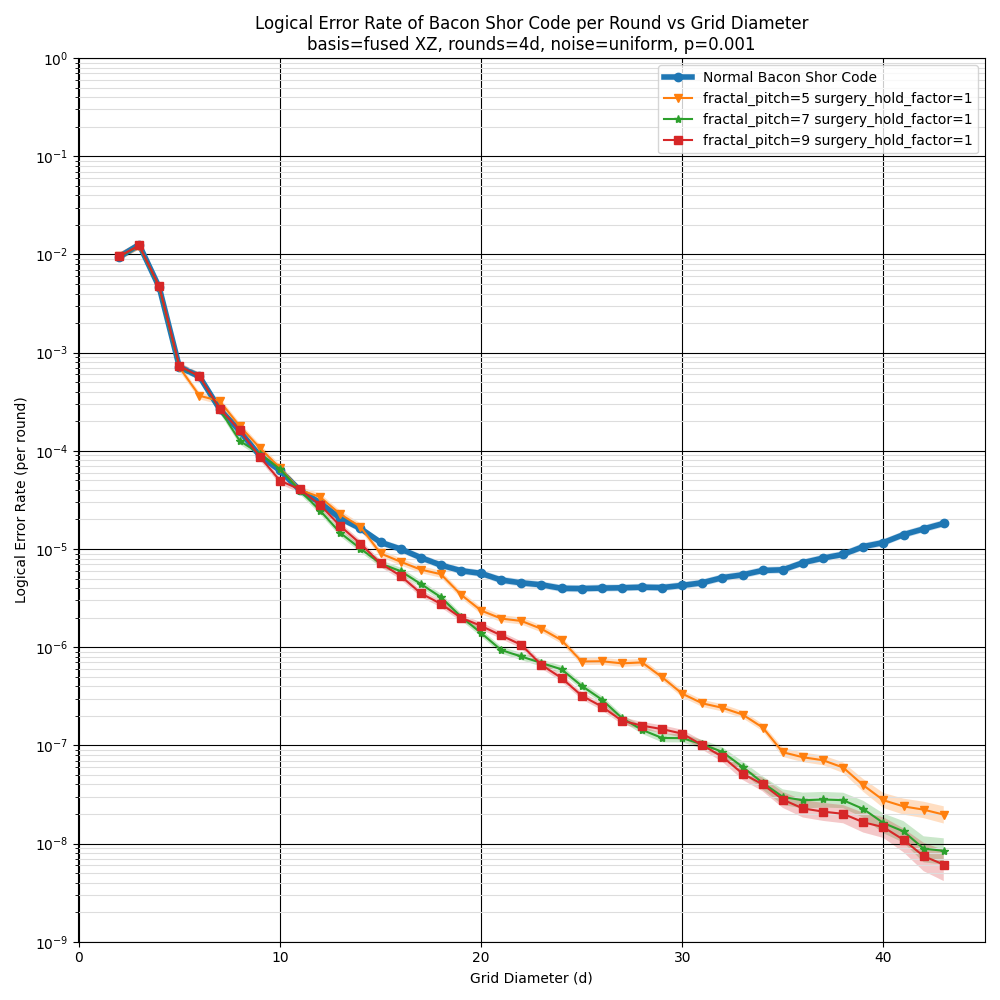}
    }
    \caption{
        Logical error rate of the normal Bacon-Shor circuit, and the fractal edge-deleting variant introduced in this paper at various ``fractal pitches''.
        The error rate is the chance of an X error and/or Z error occurring, per round, derived from \fig{error_rate_x} and \fig{error_rate_z} by assuming they are independent failures.
        The normal Bacon-Shor code fails to get below an error rate of $10^{-6}$.
        It reaches its best error rate at around a grid diameter of 25.
        The fractal variants instead continue to improve as the grid diameter increases.
        The ripples in the curves are due to the grid diameter going into and out of phase with powers of the fractal pitch, because we didn't attempt to evenly distribute the logical qubit sizes.
        Highlights correspond to hypotheses with a likelihood within a factor of 1000 of the max likelihood hypothesis, given the sampled data.
    }
    \label{fig:error_rate}
\end{figure}

Our main numerical result is that, at a noise strength of $p=0.001$, the Bacon-Shor code stops improving at a code distance of 25 while our construction is still improving at distance 40 (see \fig{error_rate}).
These simulated experimental results don't necessarily confirm that the construction has a threshold.
The ripples in the curves (as the grid diameter goes in and out of phase with powers of the fractal pitch) make it hard to tell if the improvement is consistent or slowing.
Regardless, this data clearly demonstrates that you can substantially outperform the normal Bacon-Shor code simply by deleting the right pattern of measurements.

For a proof that the fractal construction has a threshold, see \app{threshold}.
Note that the honeycomb code gets better performance from the same gate set using strictly sparser connectivity~\cite{hastings2021dynamically,paetznick2023honeycomb,gidney2022honeycombplanar,kesselring2022anyoncondense}.

Before we finish we want to note something counter-intuitive we found, which the careful reader may have noticed as strange about the $\text{Include}_{b,f}(e, t)$ function defined in the previous section.
The performance of the system was best when we held in the lattice surgery stitched configuration for 1 round, rather than multiple rounds (see \fig{error_rate_x} and \fig{error_rate_z}).
This shouldn't be true in general, since holding for 1 round breaks the fault tolerance of the logical measurement.
A single physical pair measurement error can invert a 1-round lattice surgery measurement.
We think the reason that holding for 1 round works best is specific to our use case, due to our construction \emph{repeating} logical pair measurements which are \emph{waiting} for each other.
Holding for 1 round results in less reliable measurements, but they are still being cross-checked across time, and they are happening faster.
We don't know if this 1-round-works-best effect remains true for arbitrarily high levels of concatenation, but it was the clear winner for the sizes that we simulated.

\section{Conclusion}
\label{sec:conclusion}

In this paper, we showed how to perform planar concatenation of the Bacon-Shor code by using lattice surgery.
Our construction is equivalent to deleting gates from a larger Bacon-Shor code.
This means that the performance of the Bacon-Shor code can be qualitatively improved, from boundedly good to arbitrarily good, purely by doing less.

One interesting fact about our construction is that it inherently requires time dynamics.
Pair measurements come and go in a fractal pattern, making it impossible to write down the construction as a static stabilizer code or gauge code.
This is yet another example of how adding time dynamics, or understanding time dynamics, can improve quantum error correction~\cite{hastings2021dynamically,mcewen2023relaxingsurface}.
It's also an example of the power of fractal order parameters in quantum error correction~\cite{yoshida2013exotic}.

\section{Contributions}

Both authors discussed ideas for using lattice surgery to concatenate the Bacon-Shor code under planar connectivity.
Dave Bacon attempted to write code implementing the construction.
Craig Gidney wrote code implementing the construction, and insisted on the title.

\section{Acknowledgements}

We thank Mike Newman for pointing us to previous work on the Bacon-Shor code.
We thank Austin Fowler for reviewing the paper and making suggestions that improved it.  Dave Bacon is a CIFAR Associate Fellow in the Quantum Information Science Program.
We thank the Google Quantum AI team for creating an environment where this work was possible.

\printbibliography

\clearpage
\appendix

\section{Noise Model}
\label{app:noise_model}

All circuits in this paper were simulated using the uniform noise model defined in \tbl{noise_model}.

\begin{table}[H]
    \centering
    \begin{tabular}{|r|l|}
    \hline
    Noise channel & Probability distribution of effects
    \\
    \hline
    $\text{MERR}_B(p)$ & $\begin{aligned}
        1-p &\rightarrow M_{B}
        \\
        p &\rightarrow M_{(-1 \cdot B)} \text{\;\;\;\;\;\emph{(i.e. measurement result is inverted)}}
    \end{aligned}$
    \\
    \hline
    $\text{XERR}(p)$ & $\begin{aligned}
        1-p &\rightarrow I
        \\
        p &\rightarrow X
    \end{aligned}$
    \\
    \hline
    $\text{ZERR}(p)$ & $\begin{aligned}
        1-p &\rightarrow I
        \\
        p &\rightarrow Z
    \end{aligned}$
    \\
    \hline
    $\text{DEP1}(p)$ & $\begin{aligned}
        1-p &\rightarrow I
        \\
        p/3 &\rightarrow X
        \\
        p/3 &\rightarrow Y
        \\
        p/3 &\rightarrow Z
    \end{aligned}$
    \\
    \hline
    $\text{DEP2}(p)$ & $\begin{aligned}
        1-p &\rightarrow I \otimes I
        &\;\;
        p/15 &\rightarrow I \otimes X
        &\;\;
        p/15 &\rightarrow I \otimes Y
        &\;\;
        p/15 &\rightarrow I \otimes Z
        \\
        p/15 &\rightarrow X \otimes I
        &\;\;
        p/15 &\rightarrow X \otimes X
        &\;\;
        p/15 &\rightarrow X \otimes Y
        &\;\;
        p/15 &\rightarrow X \otimes Z
        \\
        p/15 &\rightarrow Y \otimes I
        &\;\;
        p/15 &\rightarrow Y \otimes X
        &\;\;
        p/15 &\rightarrow Y \otimes Y
        &\;\;
        p/15 &\rightarrow Y \otimes Z
        \\
        p/15 &\rightarrow Z \otimes I
        &\;\;
        p/15 &\rightarrow Z \otimes X
        &\;\;
        p/15 &\rightarrow Z \otimes Y
        &\;\;
        p/15 &\rightarrow Z \otimes Z
    \end{aligned}$
    \\
    \hline
    \end{tabular}
    \caption{
        Definitions of noise channels used to define noisy versions of gates in \tbl{noise_model}.
    }
    \label{tbl:noise_channels}
\end{table}

\begin{table}[H]
    \centering
    \begin{tabular}{|r|l|}
    \hline
    Ideal gate & Noisy gate
    \\
    \hline
    $\text{Idle}$ & $\text{DEP1}(p)$ \\
    \hline
    $R_X$ & $\text{ZERR}(p) \cdot R_X$
    \\
    $R_Z$ & $\text{XERR}(p) \cdot R_Z$
    \\
    \hline
    $M_X$ & $\text{DEP1}(p) \cdot \text{MERR}_X(p)$
    \\
    $M_Z$ & $\text{DEP1}(p) \cdot \text{MERR}_Z(p)$
    \\
    \hline
    $M_{XX}$ & $\text{DEP2}(p) \cdot \text{MERR}_{XX}(p)$
    \\
    $M_{ZZ}$ & $\text{DEP2}(p) \cdot \text{MERR}_{ZZ}(p)$
    \\
    \hline
    \end{tabular}
    \caption{
        The uniform noise model used by simulations in this paper.
        \tbl{noise_channels} defines each noise channel.
    }
    \label{tbl:noise_model}
\end{table}

\clearpage
\section{Additional Data}
\label{app:additional_data}

\begin{figure}[H]
    \centering
    \resizebox{0.55\linewidth}{!}{
    \includegraphics{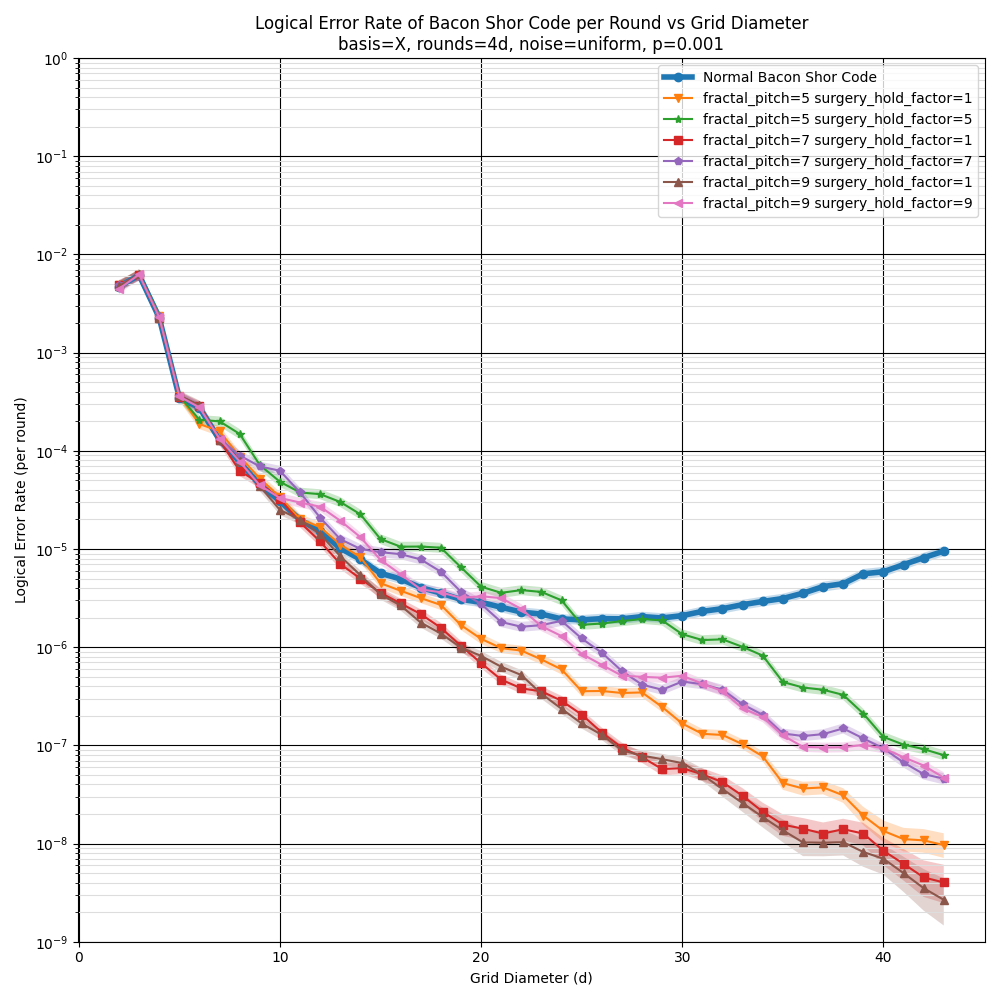}
    }
    \caption{
        Logical error rate of the X observable in Bacon-Shor circuits, at various ``fractal pitches'' and ``surgery hold factors''.
        The fractal pitch is the code distance where concatenation occurs.
        The surgery hold factor is how many times more rounds lattice surgery measurements are performed over, at each successive level of concatenation.
        We expected the best surgery hold factor to be the same as the fractal pitch, but for the parameters explored in this paper a surgery hold factor of 1 worked best.
        Highlights correspond to hypotheses with a likelihood within a factor of 1000 of the max likelihood hypothesis, given the sampled data.
    }
    \label{fig:error_rate_x}
\end{figure}

\begin{figure}[H]
    \centering
    \resizebox{0.55\linewidth}{!}{
    \includegraphics{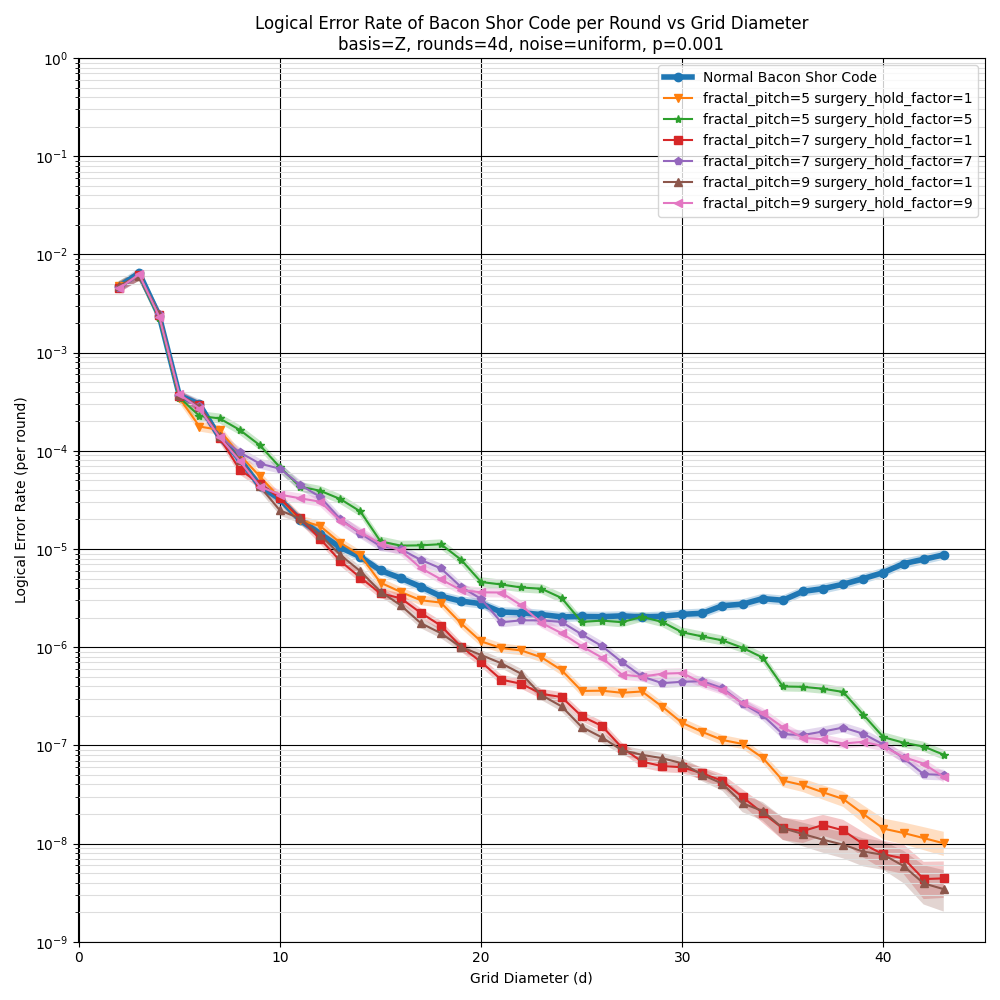}
    }
    \caption{
        Logical error rate of the Z observable in Bacon-Shor circuits.
        Very similar to \fig{error_rate_x}, due to approximate symmetries in the Bacon-Shor code and in the circuit.
        Highlights correspond to hypotheses with a likelihood within a factor of 1000 of the max likelihood hypothesis, given the sampled data.
    }
    \label{fig:error_rate_z}
\end{figure}

\begin{figure}[H]
    \centering
    \resizebox{0.55\linewidth}{!}{
    \includegraphics{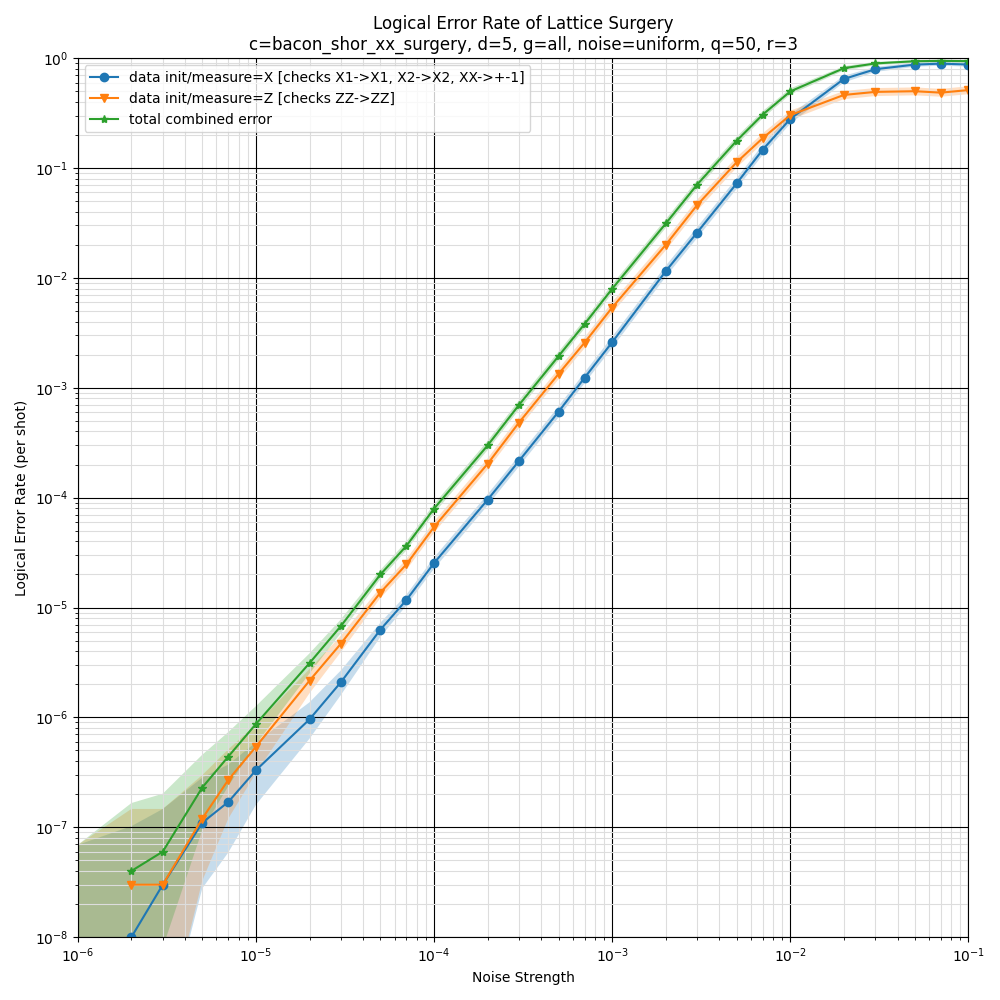}
    }
    \caption{
        Logical error rate of an XX lattice surgery measurement between two distance 5 Bacon-Shor codes, as described in \app{threshold} and \fig{lattice_surgery}.
        Performs two experiments: one with X basis transversal initialization and measurement of all qubits, and another with Z basis transversal initialization and measurement of all qubits.
        This allows all stabilizer flow generators of the pair measurement to be checked, to compute a total logical error rate on the entire operation.
        Note how the $ZZ \rightarrow ZZ$ rule fails more than all three other generators combined, due to the pair measurement temporarily elongating the Z stabilizers.
        The total logical error rate is approximately equal to $10000p^2$.
        This suggests a threshold of $10^{-4}$, which is worse than the $>10^{-3}$ suggested by \fig{error_rate}, possibly due to the use of soft decoding, or changes in the noise bias when concatenating multiple times, or various other differences in the details.
    }
    \label{fig:error_rate_xx}
\end{figure}

\section{Threshold Proof}
\label{app:threshold}

\begin{figure}[H]
    \centering
    \resizebox{0.9\linewidth}{!}{
    \includegraphics{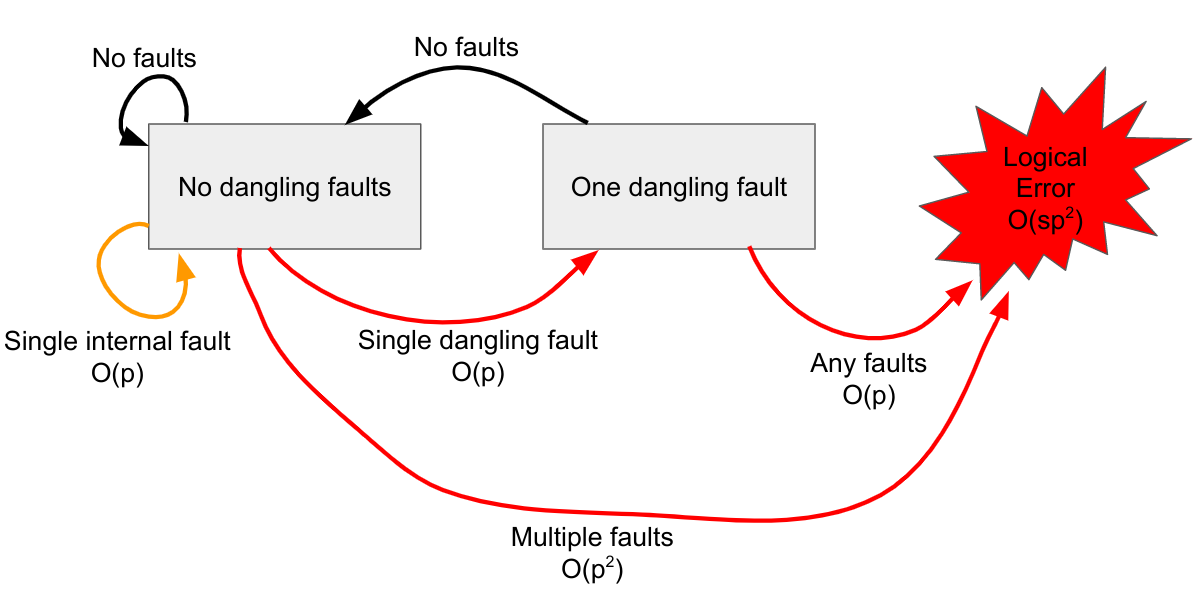}
    }
    \caption{
        A state machine that fault tolerant circuit blocks can satisfy.
        A fault is ``dangling" if it produces detection events on detectors comparing measurements at the end of one circuit block to measurements at the start of another circuit block.
        In the diagram, $p$ refers to the physical error rate and $s$ refers to the number of steps taken.
        The number of steps is analogous to the number of logical operations performed.  This diagram captures the notion of a fault not spreading disastrously for concatenated codes, as formalized in \cite{aliferis2006threshold, aliferis2011thesis}.
    }
    \label{fig:threshold_machine}
\end{figure}

We prove our construction has a threshold by showing our logical operations satisfy the state machine shown in \fig{threshold_machine}.
This state machine requires two faults to occur within one logical operation, or in two sequential logical operations, for a logical error to occur.
For the proof, we use a digitized noise model where each operation either happens correctly, or faults entirely. This model is the correct one for concatenated codes, since higher levels, unlike the lowest level, may have highly correlated faults~\cite{aliferis2006threshold}.
When an operations faults, it can arbitrarily simultaneously corrupt all the qubits it touches and all the measurement results it produces.
The gate set we're using contains single-qubit initialization ($R_X$ and $R_Z$), single-qubit measurement ($M_X$ and $M_Z$), pair measurement ($M_{XX}$ and $M_{ZZ}$), and idling ($I$) where we assume idling takes as long as a pair measurement.
We will show that the logical version of each of these operations meets the requirements of the state machine.

We divide circuits into ``rounds'', where each round is composed of 4 layers of gates and measures the parity of every active edge exactly once.
We'll use distance 5 Bacon-Shor, instead of distance 3, because when a pair measurement beaks it can corrupt two qubits and this would not be correctable at distance 3.
We'll always use enough rounds for measurement errors to be correctable.

Consider the $M_{XX}$ measurement shown in \fig{lattice_surgery}.
Suppose it spends 1 round in the unstitched configuration (``before''), then 3 rounds in the stitched configuration (``during''), then 1 round in the unstitched configuration (``after'').
This involves performing 215 $M_{XX}$ gates, 200 $M_{ZZ}$ gates, and 170 idle gates, for a total of 585 gates.
Note that these physical gates are all local (see \fig{lattice_surgery}).

By enumerating all possible failures on all possible gates during the logical $M_{XX}$, we find that any individual fault can be corrected, except for some faults on gates in the last round.
We classify faults occurring in the last round as dangling faults, which will be passed along to the next logical operation on one of the logical qubits.
When correcting the detection events within a logical operation, any detection events which may have been produced by a dangling fault are ignored and left for the next logical operation.
That operation is responsible for detecting and correcting the dangling fault.
Note that this involves comparing to the measurements at the end of this operation (the previous operation, from the perspective of the next operation) in order to detect and correct any dangling faults.
We can also check by enumeration that the $M_{XX}$ circuit has the property that it will correct any dangling faults arriving into it as input, for each possible combination of previous operations.
With this done, we've confirmed the $M_{XX}$ logical gate behaves at least as well as the state machine in \fig{threshold_machine}.

To experimentally demonstrate that the logical pair measurement can actually correct any individual error, we performed simulated Monte-Carlo sampling.
The results are shown in \fig{error_rate_xx}.
It confirms the logical error rate is quadratically suppressed compared to the physical error rate, under circuit noise.

Because the $M_{XX}$ gate is our most complicated gate, showing that it satisfies the necessary conditions is the bulk of the proof.
The analysis of the $M_{ZZ}$ gate is identical, but with $X$ replaced by $Z$.
The logical $I$, $R_X$, and $R_Z$ gates are the same idea, but on smaller chunks of circuit.
The main detail left is that the logical $M_X$ and $M_Z$ gates are not allowed to leave dangling faults.
We can show they have these properties by enumerating all possible single faults.

All of our operations behave at least as well as a state machine whose logical error is quadratic in the physical error rate and linear in the number of steps.
Recall that the edge-deletion construction presented in this paper is equivalent to concatenating the Bacon-Shor code using lattice surgery.
By concatenating, we can repeatedly quadratically suppress error.
At a sufficiently low initial error rate, this will arbitrarily reduce the error.
Therefore our construction has a threshold.

That said, something we consider lacking about this proof is that it's done in terms of concatenation, instead of in terms of the problem the decoder is actually solving in our simulations.
The decoder is not told about the concatenated structure, it's just given a single matching graph problem and decodes it like any other matching graph problem.
In particular, based on this concatenation-style proof, you wouldn't predict that holding the internal lattice surgery measurements for one round would outperform holding them for multiple rounds.
Holding for one round violates one of the conditions of the proof, but we see in simulation that holding for one round performs better.
If the proof was instead in terms of, say, percolation over the fractal structure of the matching graph, then that proof would maybe explain this otherwise surprising behavior.

\end{document}